\documentstyle[prd,aps,preprint]{revtex} 
\tightenlines 
\newcommand
{\be}{\begin{eqnarray}} 
\newcommand {\ee}{\end{eqnarray}} 
\def\be{\begin{equation}}
\def\ee{\end{equation}}

\def\ba{\begin{array}} 
\def\ea{\end{array}} 
\def\bea{\begin{eqnarray}} 
\def\eea{\end{eqnarray}}

\newcommand{\lsim}{{\;\raise0.3ex\hbox{$<$\kern-0.75em\raise-1.1ex\hbox{$\sim$}} 
\;}}
\newcommand{\gsim}{{\;\raise0.3ex\hbox{$>$\kern-0.75em\raise-1.1ex\hbox{$\sim$}}
\;}} 
\begin{document} 
\rightline{DO-TH 02/03}
\rightline{MPI-PhT/2002-05}
\rightline{UMD-PP-02-037}
\begin{center}
{\large \bf Neutrino Magnetic Moment, Large Extra Dimensions and High
Energy Cosmic Neutrino Spectra}

\medskip
{K. R. S. Balaji\footnote{balaji@zylon.physik.uni-dortmund.de}}

{\it  Institut f\"ur Theoretische Physik, Universit\"at Dortmund,
Otto-Hahn Str.4, 44221 Dortmund, Germany.}

\smallskip
{Amol S. Dighe \footnote{amol@mppmu.mpg.de}}

{\it Max-Planck-Institut f\"ur Physik, F\"ohringer Ring 6, D-80805,
M\"unchen, Germany.}

\smallskip
{R.N. Mohapatra\footnote{rmohapat@physics.umd.edu}}

{\it Department of Physics, University of Maryland, College Park,
MD-20742, U.S.A.}

\end{center}
\begin{abstract}
We point out that the presence of bulk neutrinos in models with large
extra spatial dimensions can lead to observable flavour specific deformations 
in the spectra of extreme high energy cosmic neutrinos. These deformations
are due to the spin precession of the high energy neutrinos
in the background magnetic fields via electromagnetic interactions. 
Measurements with existing and proposed neutrino telescopes which are meant 
to detect high energy neutrinos can therefore provide a novel way to probe
the size of extra hidden dimensions. We qualitatively illustrate the flavour 
suppression due to the Earth, Sun and intergalactic magnetic fields. An 
observable consequence of this precession could be an angular asymmetry 
for the extreme high energy neutrinos from the atmosphere and
flavour specific deformations of the intergalactic neutrinos. 
\end{abstract}

\newpage

\section{Introduction}
The possibility that there may be extra hidden dimensions in Nature, first
suggested by Kaluza and Klein, is an intriguing one. Recent developments in
string theories have suggested that there may be fundamental reasons for
their existence connected with the consistency and symmetry requirements
of the string world sheet. This has led to an explosive revival of
interest in the physics implications of extra dimensions
 and many investigations have been carried out in this area in the past
twenty years.

A particularly interesting framework for these investigations is the
so called brane-bulk setup where the standard model (SM) particles live in
a
3+1 dimensional brane subspace of the full multi-dimensional universe 
whereas the gravity and other SM singlet
fields reside in the entire 3+1+n dimensional space. This picture arises
in the nonperturbative D-brane solutions of string theories and is thus
believed to have solid theoretical foundation. 

This picture allows a very novel approach to gravity according to which the 
Planck scale is not a fundamental scale in Nature but rather a function of the
fundamental variable, the string scale $M_*$. The ``radii of
compactification'' $R_i$ of the extra hidden dimensions and the familiar 
Planck scale $M_{P\ell}$ are related by
\be
M^2_{P\ell}= M^{2+n}_* (2\pi)^n R_1 R_2 R_3 \cdot\cdot\cdot R_n~.
\ee
The scenario $n=1$, with all but one $R_i$ of order $M_*$ is ruled out for
$M_* \sim$ TeV by the observed
accuracy of Newton's laws over the Earth-Sun distances.
If only two extra dimensions are significantly larger than $M_*^{-1}$,
i.e. $R_1\simeq R_2=R$ and $R_3\simeq R_4\simeq \cdot\cdot R_n= M^{-1}_*$, 
we get $R\simeq$ millimeter for $M_*\simeq $ TeV. This scenario
is allowed by the current data, since the inverse square law for
gravity has not been tested for distances less than mm.
Since the small radii $R_3,R_4$ have no effect on the physics, 
this is equivalent to a scenario with $n=2$.
Alternatively, one may envision one large extra dimension with 
$R_1 \sim$ mm, with the rest much smaller e.g. $r \lsim 10^{-2}$ fm 
for $M_*\simeq 10$ TeV. The influence of such small dimensions cannot 
be felt at any of the current  experiments, and hence this scenario of
only one large extra dimension cannot be ruled out.
This will be equivalent to a $n=1$ scenario.

A basic difference between the two scenarios described above
(n=2 and n=1) is that the Kaluza-Klein (KK) excitations of any
particle in the bulk in the first scenario are closely spaced in two extra
dimensions whereas in the second class of models it happens only in one
dimension, making the phase space structure very different. 
The two scenarios then have different phenomenological
consequences, and it is important to distinguish between them
through their experimentally testable consequences. Perhaps more
important than this is to test the possibility of large extra
dimensions itself. The
goal of this paper is to provide one such test in the domain of neutrinos.

The small mass of the neutrinos, for which there is now very strong
evidence from solar and atmospheric neutrino data, is understood in
these models in a way which is very different from the conventional 3+1 
dimensional unified theories. The simplest possibility is to postulate the 
existence of SM singlet neutrinos in the bulk and couple 
them to the SM neutrinos in a gauge invariant way using the usual Higgs
mechanism. This leads to naturally small masses for the SM neutrinos
\cite{dienes} due to small overlap of the bulk neutrino wave function on
the brane.
The coupling of the SM neutrinos with the bulk neutrino provides
a naturally massless sterile neutrino and a tower of KK neutrinos
that also mix with the SM neutrinos. 
This leads to a variety of interesting tests and
constraints on the brane bulk models; for instance for tests
using solar neutrinos, see \cite{dvali} and tests using high energy muons,
see \cite{tetradis}.

There are many possible scenarios for neutrino masses in brane-bulk
models.  A common feature of all those models is the presence of
a SM singlet neutrino mentioned above. Regardless of the
details of the model, as long as the known neutrinos mix with the KK modes
of the bulk neutrino, there is a magnetic moment connecting the brane
neutrinos with the bulk $\nu_B$ modes \cite{ng}. 
 We will show below that
this can give rise to observable deformation of the high energy neutrino 
spectra. The flavour composition of the high energy neutrinos may also
change; the final flavour content will depend on the neutrino 
mass hierarchy. This can provide 
valuable information about the extra dimensions, and possibly also
about the hierarchy structure of the neutrino masses.
These spectral deformations occur over a wide range of energies starting
from one TeV to 1000 TeV and higher. The neutrino telescope facilities
such as ANTARES and ICE CUBE, which can look for these extreme high energy
muon and tau neutrinos can therefore provide valuable information on
models with large extra dimensions.\\

 Our paper is organised as follows. In the next section, we describe the 
basic framework of the model. We illustrate the enhancement of the neutrino
magnetic moment with energies and give the expected strengths depending on 
the neutrino mass hierarchy. We then apply this result in section III for
ultra-high energy neutrino spectrum and examine the changes in the individual
neutrino fluxes. Although, these changes are dependent on the mass hierarchy 
they are independent of neutrino oscillations. We discuss
the physical situations under which we can have flavour suppression and in 
particular, we focus on the spin precession effects due to the Earth, Sun and
intergalactic magnetic fields. In section IV, we describe a useful 
angular asymmetry which can probe the effects of neutrino magnetic moments in 
extra-dimension scenarios. Finally, we conclude in section V by summarising 
the main results.

\section{Minimal Scenario}
\label{minimal}

We will work with the simplest scenario where there are only three extra 
neutrinos in the bulk and the usual three generations of fermions of the 
SM residing in the brane. The neutrino mass in these models arises from 
the brane bulk Yukawa interactions of the type
\be
{\cal L}~=~\frac{h_{\alpha \beta}}{{M_*}^{n/2}}\int dy ~\delta(y)~
\bar{L}_\alpha H \nu_{B, \beta}(x,y) +
h.c.~.
\ee
 If the bulk neutrino resides in (3+1+n) dimensions, where
$\delta(y)$ is a n-dimensional delta function and $dy$ is the
n-dimensional measure in the hidden space. Taking the individual modes on
the brane located at $y=0$, after symmetry breaking  one gets a Dirac 
mass matrix for the active neutrinos given by
\be
m_{D, \alpha \beta}=\frac{h_{\alpha\beta}vM_*}{M_{P\ell}}~ \sim
6 \times 10^{-5} ~{\rm eV}~ h_{\alpha\beta} 
\frac{M_*}{{\rm TeV}}~~.
\ee

These neutrinos also mix with the KK modes of the bulk neutrinos.
Diagonalization of this mass matrix leads to three eigenvalues $m_{i}$
and a mixing matrix $U_{\alpha i}$ that characterizes the
neutrino oscillations.
The mixing with the bulk fermion modes is characterized 
 for the case of one dimension
by three parameters $\xi_i\simeq \sqrt{2}m_{i}R$. 
 This minimal model induces a
magnetic moment connecting $\nu_i$ from the brane to each of the bulk
modes $\nu^{(k)}_B$, in addition to that  given by the SM result \cite{shrock}
\be
\mu_i \simeq 1.6 \times 10^{-19} \mu_B ~\frac{\sqrt{2}m_{i}}{1~eV}~,
\ee
where $\mu_B$ is the Bohr magneton.
Note that the value of the magnetic moment is connected to the mass of the
neutrino and through it to the observed neutrino oscillation parameters.
We will exploit this connection in our search for extra dimensions using
high energy neutrinos from the cosmic rays.

Before getting to the discussion of realistic extra dimensional scenarios 
for
neutrinos, let us first outline the effect of the magnetic moment on a
high energy neutrino travelling through a magnetic field.
If the neutrino energy is denoted by $E$, then all KK modes of the
bulk neutrino upto energy $E$ will be excited by the magnetic moment
interaction. This enhances the effect of the spin flip due to magnetic
moment over the case with no large extra dimension as in SM. In the generic 
case, the effective magnetic moment becomes \cite{ng}
\be
\tilde\mu \approx 4hA_n\left( \frac{1}{2\pi} \frac{E}{M_*} \right)^{n/2}
\times 10^{-8} \mu_B~.
\label{mueff}
\ee

Here, $h$ is the strength of the brane-bulk Yukawa coupling,
 and $A_n$ is the volume of the
positive hemisphere of a n dimensional unit sphere.
Note that $\tilde\mu$ is inversely proportional to the value of
the fundamental scale $M_*$, whereas with a given value of the
scale, it grows with the energy of neutrino. This last feature 
provides us with the possibility of having large 
effective magnetic moments for neutrinos at high energies, while
still obeying the constraints provided by the experiments at 
lower energies. 

The current strongest limits on the neutrino magnetic moment are
$\mu_\nu \leq 1.6 \times 10^{-10} \mu_B$ at around 10 MeV 
\cite{reac-lim} from reactor experiments and SK solar neutrino data and
$\mu_\nu \leq 10^{-12} \mu_B$ from supernova 1987a \cite{astr-lim},
for neutrino energies of 10-50 MeV.
Note that these limits now get translated to limits on the $\tilde\mu$,
which depends on two parameters $(h, M_*)$. So clearly, the limits can be
satisfied by appropriate choice of $h$ and would therefore fail to provide
any useful information on the string scale $M_*$. One may however hope to
combine other terrestrial data to extract useful information.

A further remark is in order.
If $E$ is higher than the string scale, one would excite
the stringy modes rather than the KK modes, whose spacing is in general
higher than that of the KK modes (expected to be of order $\sim M_*$). As
a result, for energies higher than the string scale, it is probably
reasonable to use the $M_*$ as a cut-off on the phase space integral. In
any case, as we will see below, for higher dimensional models, we expect 
$M_* \sim 500-1000$ TeV, which covers almost the
entire expected range of neutrino energies (upto $10^{15}$ eV or so). We
will assume in
all our considerations that it is only the KK modes that are important.
In fact, as the neutrino energy gets higher, our method probes higher
values of the fundamental scale $M_*$.
 In this paper, we will focus on the effects due to $M_* \sim 10-10^3$ Tev
and hence neutrino energies in the intermediate range, 
$10^{10} \mbox{eV} \leq E \leq 10^{15}\mbox{eV}$.\\

To get a feeling for the kind of $\tilde\mu$ one can expect in realistic
examples consistent with known data, we will consider  
constraints coming from attempts to fit the
observed solar and atmospheric neutrino oscillation
data \cite{perez,davu,cremi}, although this does not account
for the LSND anomaly.
Since the expression for $\tilde\mu$ involves the Yukawa coupling
$h$ and the string scale $M_*$ which are connected to the value of neutrino
masses,  acceptable neutrino mass patterns from oscillation data will
also restrict the allowed values of $\tilde\mu$.
First important point to recall is that in the minimal scenario described here 
the LSND observations cannot be accomodated \cite{perez,davu}.
However, it is possible to find a domain of parameter space in
$h$ and $M_*$ that explains the neutrino data and does not violate
the experimental constraints on $\tilde\mu$ mentioned above. Indeed, in the
limits $h \to 0$ or $M_* \to \infty$, the oscillations between
active neutrinos account for the solar and atmospheric anomalies
successfully.\\

Since the  bulk modes are sterile neutrinos, the recent SNO data on solar
neutrinos and the Super-Kamiokande data on atmospheric neutrinos can be
understood only if we choose the parameter region $\xi_i \ll 1$.
The mixings are then dominated by the $U_{\alpha i}$ discussed in 
\cite{perez,davu}. The individual neutrino
masses can have one of the following patterns \cite{cald}:\\

(i) Normal hierarchy:  $m_1 \ll m_2 \ll m_3 $~;

(ii) Inverted hierarchy $m_1 \sim m_2 \gg m_3 $~;

(iii) Degenerate: $ m_1 \simeq m_2 \simeq m_3$ .\\

Here the neutrino mass eigenstates are defined such that the familiar 
atmospheric $(\Delta m^2_A)$ and solar $(\Delta m^2_\odot)$ mass-squared
differences, i.e.
$|\Delta m^2_{31}| \approx |\Delta m^2_{32}| \approx \Delta m^2_{A} 
\approx 3 \times 10^{-3}$ eV$^2$, and $|\Delta m^2_{21}| \approx
\Delta m^2_\odot$.\\

Since the magnetic moment is directly given by the parameters $m_i$, the
effective magnetic moments follow the same pattern as the mass hierarchies.
Therefore, we can write the contribution due to the $i^{th}$ mass eigenstate 
as
\be
\tilde\mu_i\approx 7 \left( \frac{1}{2 \pi}
\frac{E}{M_*} \right)^{n/2} \left( \frac{{\rm TeV}}{M_*} \right)
\left( \frac{m_i}{{\rm eV}} \right)\times 10^{-4} \mu_B ~.
\label{mu-mi}
\ee

For instance, in case (i), we have  $\tilde \mu_1 \ll \tilde\mu_2 \ll
\tilde \mu_3$. In the case of normal hierarchy,
if we choose the maximum value of the Yukawa coupling $h$
to be one, then atmospheric neutrino data implies
$m_3 \approx 0.06$ eV, and hence $M_* \sim  10^{3}$ TeV.
Using (\ref{mu-mi}), we then get
 
\bea
\tilde\mu_3 & \approx & 5 \times 10^{-10} \mu_B (E/TeV)^{1/2} 
\quad {\rm  (n=1)} ~,\\
\tilde\mu_3 & \approx & 7 \times 10^{-12} \mu_B (E/TeV) 
\quad {\rm (n=2)} ~.
\eea

Since the LMA solution to the solar neutrino data 
requires $m_2\simeq 3 \times 10^{-3}$ eV, we have 
\bea
\tilde\mu_2 & \approx & 2.5 \times 10^{-11} \mu_B (E/TeV)^{1/2} 
\quad {\rm  (n=1)} ~,\\
\tilde\mu_2 & \approx & 3.5 \times 10^{-13} \mu_B (E/TeV) 
\quad {\rm (n=2)} ~.
\eea

The value of  $\tilde\mu_1$ is much smaller. Since the spin-flavour
precession is proportional to $\tilde\mu_i$ the precession phase will
be much smaller for $\nu_1$ and $\nu_2$ as compared to that for $\nu_3$.
Similar observations can be made about the other mass hierarchies which we 
shall discuss below.

\section{Magnetic moment effects on intergalactic high energy neutrinos}

Here, we consider the effects of the induced magnetic moment on high 
energy neutrinos in the presence of galactic magnetic fields through which
they will travel, and as a result, may spin-precess to the bulk
states. Depending on the amount of precession, this may lead to a
depletion of various neutrino species. We assume that
the precession phase gained by a mass eigenstate $\nu_i$ is
\be
\phi_i(E) = \int  \tilde\mu_i(E) B . dl~,
\ee
where $dl$ is the distance travelled through a magnetic field $B$.
Physically, this is the phase gained in the mass basis, due to a transition 
from the active left-handed modes to the sterile right-handed bulk modes. 
The corresponding precession probability for the $i^{\rm th}$ mass eigenstate 
is then
\be
P_i(E) = \sin^2 \phi_i(E)~.
\ee

It is important to note that there are two simultaneous effects here: one
due to the flavour mixing and another due to magnetic moment precession.
The final flavor content at the detector therefore depends on the distance
travelled and the energy. If an extremely long distance of travel is 
involved, two mass eigenstates with a mass squared difference, 
$\Delta m^2_{ij}$ can lose their coherence over a distance scale of 
$E/\Delta m^2_{ij}$ and travel as independent mass eigenstates. The 
coherence length scale is
\be
L_c \sim 10^8 {\rm cm } \left(\frac{E}{TeV}\right) 
\left(\frac{{\rm eV}^2}{\Delta m^2_{ij}}\right)~~.
\label{coh}
\ee
Loss of coherence happens specifically when the distances involved
are intergalactic distances. An example of this are neutrinos which are 
produced through sources like AGNs. The final flavour content will depend 
on whether coherence is lost or it remains. In this section, we will
focus on intergalactic neutrinos, which are likely to lose coherence.\\

The neutrinos from cosmic sources are produced mainly from the decays of 
pions. This implies that the ratios of the production fluxes of the electron, 
muon and tau neutrinos are
\be
F_e^{(0)} : F_\mu^{(0)} : F_\tau^ {(0)} \approx 1:2:0~~.
\ee
As long as the mixing between $\nu_\mu$ and $\nu_\tau$ is maximal
(which is strongly supported by the atmospheric neutrino data at SK), the
three mass eigenstates have the production fluxes
\be
F_1^{(0)} : F_2^{(0)} : F_3^ {(0)} \approx 1:1:1~~.
\ee
As already mentioned, due to loss of coherence, the three mass
eigenstates travel independently and if
there is no magnetic moment induced precession, the fluxes of
different flavours at the detector are determined by flavour 
mixings and are given by 
\be
F_\alpha = \sum |U_{\alpha i}|^2 F_i~~.
\label{prob}
\ee
For bimaximal mixing case, the flavour composition 
of the neutrino flux at the detector will be \cite{athar}
\be
F_e : F_\mu : F_\tau \approx 1:1:1 ~~.
\label{flux}
\ee
The spin precession will modify this relation, since
these neutrinos travel a distance of several $\sim$ Mpc, even though the
magnetic
fields along the path are small ($B \lsim 10^{-8}$ Gauss). We estimate the 
precession phase by assuming (for example) that the neutrinos encounter a
coherent  magnetic field of this magnitude for a typcial distance of
Mpc. This gives
 
\be
\phi_i \sim 10^{13} (\tilde\mu_i/\mu_B)~.
\label{galac}
\ee
Since $\tilde{\mu}_i$ depends on the neutrino mass, the phase $\phi_i$ 
that a particular mass eigenstate precesses through can be large if the
corresponding neutrino mass is ``large'', and averaging for large
phases would give
$\langle \sin^2 \phi_i \rangle \approx 0.5$, and there will be a
suppression of the flux $F_i$ by a factor of two. If a particular mass is
small, there will be no magnetic precession effect and no additional 
suppression. The resulting flavour composition of the detected neutrinos 
for different mass patterns of the neutrinos is given below.\\

(i) For the normal mass hierarchy, $\tilde\mu_1 \ll \tilde\mu_2 \ll
\tilde\mu_3$. Hence, it is possible for $\phi_3 \sim 1$,
whereas $\phi_1 \ll \phi_2 \ll 1$. In this case, we get
$$F_1:F_2:F_3 = 1:1:0.5 \quad {\rm or} \quad  
F_e:F_\mu:F_\tau = 1:0.75:0.75~.$$

If $\phi_3 \gg 1, \phi_2 \sim 1$ and $\phi_1 \ll 1$, we have for 
bimaximal mixing
$$F_1:F_2:F_3 = 1:0.5:0.5 \quad {\rm or} \quad  
F_e:F_\mu:F_\tau = 0.75:0.625:0.625~~.$$

If on the other hand, the mixing of $\nu_e$ with the other states
is not large (this possibility is disfavoured, but not yet
completely ruled out), we get
$$F_1:F_2:F_3 = 1:0.5:0.5 \quad {\rm or} \quad  
F_e:F_\mu:F_\tau = 1:0.5:0.5~~. $$
If the precession of all three mass eigenstates is large, all
the flavours are suppressed by a factor of two, i.e.
$$F_1:F_2:F_3 = 0.5:0.5:0.5 \quad {\rm or} \quad  
F_e:F_\mu:F_\tau = 0.5:0.5:0.5~~. $$

(ii) For the inverted mass hierarchy, if $\phi_1 \approx \phi_2 \sim 1$ and 
$\phi_3 \ll 1$, then
$$F_1:F_2:F_3 = 0.5:0.5:1 \quad {\rm or} \quad  
F_e:F_\mu:F_\tau = 0.5:0.75:0.75~~. $$

The second possibility is that the precession of all three mass eigenstates 
is large, in which case we get a uniform suppression of a factor of two
as shown above.\\ 

(iii) If the masses of all three mass eigenstates are degenerate, 
then either all mass eigenstate undergo large precession giving an
overall suppression by a factor of two, or none of the
eigenstates undergo significant precession, so that the
fluxes remain umchanged.\\

Note that in the case of inverted hierarchy and degenerate neutrino
scenarios, the final fluxes do not depend on the extent of mixing
of the electron flavour in $\nu_2$. These flavour specific deformations 
are to be contrasted with the conventional expectations in (\ref{flux}).
Current and planned neutrino telescope experiments, which are going to 
measure the flux of high energy muon and tau type neutrinos 
(the latter by the so called double bang events) could be sensitive to these
changes. Such measurements can be compared with theoretical predictions of 
the fluxes to draw conclusions about the extra dimensions.

One possible drawback regarding the above conclusion is that no reliable
information is available about the intergalactic 
magnetic fields. For instance, if these magnetic 
fields are incoherent with a spatial spread shorter than the coherent length
for spin precession, then the average spin precession phase gained by a  
neutrino mass eigenstate, $\langle \phi_i\rangle$, could be smaller. For
instance, if the source distance is $D$ and the coherence length of the
magnetic field is $\ell_c$, then the total length in the Eq. (11) can 
be replaced in the random walk approximation by $\sqrt{D \ell_c}$.
This could lead to nontrivial effects for higher energy extragalactic
neutrinos.

 Furthermore, in the physical situation considered, we require 
neutrinos from cosmologically distant sources, wherein, the 
production mechanisms are source dependent and yet unclear
\cite{cosmic}. 
This motivates us to analyse the effects due to the Earth's magnetic field
on neutrinos produced in the atmosphere, which we discuss in the next
section.

\section{Effects on extreme high energy atmospheric neutrinos}

In this section, we focus on neutrinos with energies below
$E < 10^{15}$ eV produced in the atmosphere. The total neutrino spectrum
in this energy range is
expected to be dominated by the atmospheric neutrinos\cite{mann} that
originate from the
decays of pions that are generated from cosmic rays in Earth's 
atmosphere. The spectrum of neutrinos then has the same shape
as the cosmic ray spectrum, which has been well measured in this
energy range and has a simple power law dependence. The magnitude
of the flux may be fixed by extrapolating the flux of the atmospheric
neutrino spectrum, which is measured for energies $\sim$ GeV,
with the same power law.
Using the energy range $E < 10^{15}$ eV has one more advantage --
as mentioned before, this energy would be below the fundamental scale
of the theory, so that the interactions with the stringy modes
may be neglected. In the following, we address the spin precession of these
high energy neutrinos due to the magnetic field of the Earth.\\

The neutrinos that are produced in the atmosphere of the Earth travel 
through the magnetic field of the Earth, which is $\sim$ 1 gauss.
If the distance travelled by these neutrinos before reaching the
detector is $L$ km, using (\ref{mu-mi}), we get
\be
\phi_i \sim 30 L(\mbox{km}) \left( \frac{\tilde\mu_i}{\mu_B}
\right)\left(\frac{m_{\nu}}{eV}\right)~.
\label{atm}
\ee
For $m_{\nu}\sim 0.06$ eV, $M_* = ~10 $ TeV, (6) implies
 that significant precession may be observed
only for neutrinos travelling distances of $\sim 10^4$ km
(through the earth). This will result in a zenith angle dependence which
we discuss here. The ``neutrino telescope'' experiments like 
ICE CUBE \cite{amanda} and ANTARES \cite{antares} are tuned to see
neutrinos 
arriving through the Earth and can therefore test for this effect.\\

We show that magnetic precession of the kind we are discussing leads to a 
novel feature of a flavour specific nonzero up-down asymmetry for these 
neutrinos. To make this effect quantitative, let us define the angular 
asymmetry in the standard form for a neutrino, $\nu_\alpha$ as

\be
A_\alpha(\theta_1,\theta_2) = \frac{R_\alpha(\theta_1)-R_\alpha(\theta_2)}
{R_\alpha(\theta_1)+R_\alpha(\theta_2)}~,  
\label{asym}
\ee

where $\theta_i$ is the azimuthal
angle. The
asymmetry owes its origin to the different distances travelled by the
neutrinos coming from the top and bottom of a Earth located detector
as we explain below.
 In (\ref{asym}), $R_\alpha(\theta_i)$ is the rate of 
neutrinos $\nu_\alpha$ in a given direction $\cos\theta_i$. 
We will focus on the high-energy neutrinos (with $E\leq 10^{15}$ eV)
a large fraction of which are induced by atmospheric air showers at a
distance of a few tens of 
kilometers above the Earth. Qualitatively, if we examine 
(\ref{atm}), we see that $\phi_i \sim O(1)$ 
only for $L \geq 10^{4}$Km. This implies that for neutrinos traversing the
entire diameter of the Earth, spin precession can lead to an up-down 
asymmetry 

\be
A^{UD}_\alpha \equiv A_\alpha(\pi,0)=
\frac{R_\alpha(\theta_1=\pi)-R_\alpha(\theta_2=0)}
{R_\alpha(\theta_1=\pi)+R_\alpha(\theta_2=0)}~. 
\label{asym1}
\ee 

The sign of this asymmetry is negative for $\nu_e$ and $\nu_\mu$ whereas
for  $\nu_\tau$, since $R_\tau(\theta_2=0) =0$ we will have a unit positive
value. In the following, we illustrate the expected flavour asymmetry for 
the three mass patterns. Note that since in this case the distance of travel 
is less than the coherence length $L_c$, the expected flavour fluxes are 
different from (\ref{flux}). Since there is not enough time/distance for 
flavour oscillation to have taken place, the original 
flavour ratio remains $1:2:0$ rather than  $1:1:1$, which is for the 
intergalactic neutrinos. To derive the up-down asymmetry, note that the 
flavour and mass eigenstates are related through a unitary mixing matrix 
$U$ by
\be
\nu_\alpha = U_{\alpha i} \nu_i~.
\label{fp}
\ee

Now, the mass eigenstates precess to the right-handed bulk modes 
$(\nu_i^s)$ in the back ground magnetic field and at the detector, we have 
for these states the relation

\be
\nu_\alpha^d = U_{\alpha i}\nu_i^d~;~
\nu_i^d = \nu_i \cos\phi_i + \nu_i^s \sin\phi_i~.
\label{fd}
\ee

 Therefore, the flavour fluxes at the detector, are easily seen to be

\be
F_\alpha^d = P_{\alpha\beta}F_\beta~\mbox{with}~
P_{\alpha\beta}= |U_{\alpha 1}U_{\beta 1}^* \cos\phi_1
+ U_{\alpha 2}U_{\beta 2}^* \cos\phi_2 + 
U_{\alpha 3}U_{\beta 3}^* \cos\phi_3|^2~,
\label{prob1}
\ee
where $F_{\beta}$ denotes the initial flux at the production point which is
taken to be a few tens of kilometers above the Earth.
Using (\ref{prob1}) and taking bimaximal mixing, we calculate the expected 
flavour fluxes at the detector for the three mass hierarchies. In doing
these calculations, we have taken the averaging
$\langle\cos\phi_i\rangle=0$ 
and $\langle\cos^2\phi_i\rangle=0.5$.\\

 The resulting values for $F_\alpha^d$ and $A_{\alpha\beta}^{UD}$ for the
three different mass spectra are:\\

(i) For the normal mass hierarchy:
$$F_e^d:F_\mu^d:F_\tau^d = 1:0.75:0.75~.$$
$$A_e^{UD}:A_\mu^{UD}:A_\tau^{UD} = 0:-0.45:1.0~.$$ 

(ii) For the inverted mass hierarchy:
$$F_e^d:F_\mu^d:F_\tau^d = 0.50:0.75:0.75~.$$
$$A_e^{UD}:A_\mu^{UD}:A_\tau^{UD} = -0.33:-0.45:1.0~.$$

(iii) For the degenerate mass hierarchy:
$$F_e^d:F_\mu^d:F_\tau^d = 0.5: 1.0 : 0~.$$
$$A_e^{UD}:A_\mu^{UD}:A_\tau^{UD} = -0.33:-0.33 :0~.$$

It is important to note that for energies close $10^{15}$ eV and higher,
the neutrino weak interaction cross section matter in the Earth becomes
important enough so that the spectrum gets suppressed for the upward going 
neutrinos\cite{rocky}. But our effect can manifest also at lower energies.

\section{Comments and Conclusion}

There can also be other possible sources of spin precession for 
neutrinos traversing through space. For example, the precession effects 
may be large if we consider the 
neutrinos that travel close to the Sun on their way to the detector. These
will be
affected by the solar magnetic field, which is as much as 
$B_\odot \approx 10$ Gauss at the surface of the Sun. If we assume the 
spatial profile of the magnetic field to be given by 
an approximate power law ($B \sim r^{-\alpha})$, where 
$r$ is the distance from the Sun, and $\alpha \sim 1.25$, the 
precession undergone by a neutrino travelling from the Sun to the Earth would 
then be 
\bea
\phi_i & \sim & \tilde\mu_i \int_{R_\odot}^{r_{earth}} dr ~B_\odot 
\left( \frac{r_\odot}{r} \right)^\alpha \approx
\tilde\mu_i\frac{B_\odot R_\odot}{\alpha -1} \\
& \sim & 10^{11} \left( \frac{\tilde\mu_i}{\mu_B} \right) ~~,
\label{sun}
\eea
The precession effects are likely to be large for
one or more mass eigenstates. In principle, this could also lead to an
angular asymmetry.\\

In summary, we have pointed out a novel flavour specific deformation of the
extreme high energy cosmic neutrino spectra in the presence of large extra
dimensions with low fundamental scale of particle interactions. We
have specifically noted how extreme high energy atmospheric
neutrinos detected on Earth would show an up-down asymmetry, unique in 
their origin due to the extra dimensional effect. The extreme high
energy galactic neutrinos are also likely to show flavour specific
deformations of their spectra. In contrast to the neutrinos produced in the
Earth's atmosphere, these galactic neutrinos are expected to show a
uniform
spherically symmetric flavour deformation. But, given the low rates of such 
high energy neutrinos, and the source uncertainties, we emphasise that the 
up-down asymmetry in (\ref{asym1}) for atmospheric neutrinos with energies 
$E\leq 10^{15}$ eV will be statistically favoured. 
The upcoming neutrino telescope experiments 
such as ICE CUBE, NESTOR\cite{nestor} and ANTARES. are 
particularly suited to detect 
these kinds of effects. An advantage of this method is that it can probe the
fundamental scales much higher than the conventional collider methods 
discussed in the literature since the magnetic moments depend on both the 
fundamental scale $M_*$ and neutrino energy $E$.

\begin{center}
{\bf Acknowledgements}
\end{center}

   The work of  Balaji is supported by Bundesministerium f\"ur Bildung, 
Wissenschaft, Forschung und Technologie, Bonn under contract no. 05HT1PEA9.
The work of R.N.M. is supported by a grant from the National Science
Foundation under grant no. PHY-0099544. A.S.D. wishes to thank D. Semikoz and 
M. Kachelriess for helpful discussions. We also thank J. Beacom and
J. Learned for comments.

\end{document}